\documentclass[
reprint,
 amsmath,amssymb,
aps,
prl,
]{revtex4-1}
\DeclareMathOperator{\Tr}{Tr}
\usepackage{graphicx}% Include figure files
\usepackage{dcolumn}% Align table columns on decimal point
\usepackage{bm}% bold math
\usepackage[colorlinks = true,
linkcolor = blue,
urlcolor  = blue,
citecolor = blue,
anchorcolor = blue]{hyperref}% add hypertext capabilities
\usepackage{comment}
\begin{document}

\preprint{APS/123-QED}

\title{Quantum hard disks on a lattice}% Force line breaks with \\

\author{Vighnesh Dattatraya Naik}
\author{Fabian Ballar Trigueros}
\author{Markus Heyl}
 % \email{}
\affiliation{%
 Theoretical Physics III, Center for Electronic Correlations and Magnetism,
Institute of Physics, University of Augsburg, 86135 Augsburg, Germany}%

\date{\today}% It is always \today, today,

\begin{abstract}
We formulate a quantum version of the hard-disk problem on lattices, which exhibits a natural realization in systems of Rydberg atoms. We find that quantum hard disks exihibit unique dynamical quantum features. In 1D, the crystal melting process displays ballistic behavior as opposed to classical sub-diffusion. For 2D, crystal structures remain intact against most defects, whereas classically they are washed out completely. We link this peculiar quantum behavior to quantum many-body scars. Our study highlights the potential of constrained 2D quantum matter to display unique dynamical behaviors.
\end{abstract}

\maketitle

%\tableofcontents

\textit{Introduction.—}The recent tremendous advances in quantum simulators have led to unprecedented experimental control of the real-time dynamics in constrained quantum matter.
This includes constraints generated by symmetries such as dipole conservation relating to fractonic systems~\cite{GuardadoSanchez2020,Scherg2021,Kohlert2023} or strong local interactions~\cite{Bernien_2017,Keesling2019,deLeseleuc2019,Browaeys_2020,Ebadi2021,Semeghini2021,Scholl_2021,Bluvstein2021}, which can lead to the emergence of quantum many-body scars~\cite{Bernien_2017,Turner2018,Serbyn2021,Bluvstein2021}.
The latter case nowadays finds a natural realization in Rydberg atom arrays due to the famous Rydberg blockade mechanism originating from the strong underlying Rydberg interactions~\cite{Browaeys_2020}.
It is of key importance to recognize that Rydberg blockade, on a general level, is equivalent to associating with each excitation an excluded volume, which on the classical level represents a paradigmatic class of systems with unique properties, such as in the context of the hard-disk problem~\cite{Li2022}.
In the quantum domain, the equilibrium properties of hard disks in the continuum have been already investigated ~\cite{Huang1957, Luban1965, Lieb1967,winer2023}.
However, the exploration of quantum hard disks on a lattice and their associated dynamical properties has, until now, remained unexplored.
In this work, we introduce a quantum hard-disk model on lattices inspired by current experiments in Rydberg atom arrays.
We show that the static properties at high temperatures are identical to the classical analogue yielding crystalline phases at sufficiently high particle densities.
However, we find that the dynamical features turn out to be fundamentally different, which we illustrate with a crystal melting process in one spatial dimension (1D) and the dynamics of crystal defects in two-dimensional (2D) quantum hard-disk systems.
In particular, for the 2D case we observe that, generically, the crystals remain intact in the presence of defects, while they are washed out on the classical level.
We associate this unique quantum feature with the presence of quantum many-body scars and Hilbert-space fragmentation.
We discuss the feasibility of implementing the quantum hard-disk problem experimentally in Rydberg atom arrays as the natural platform to realize excluded volumes through the Rydberg blockade mechanism.

\begin{figure*}
    \centering
    \includegraphics[width=\linewidth]{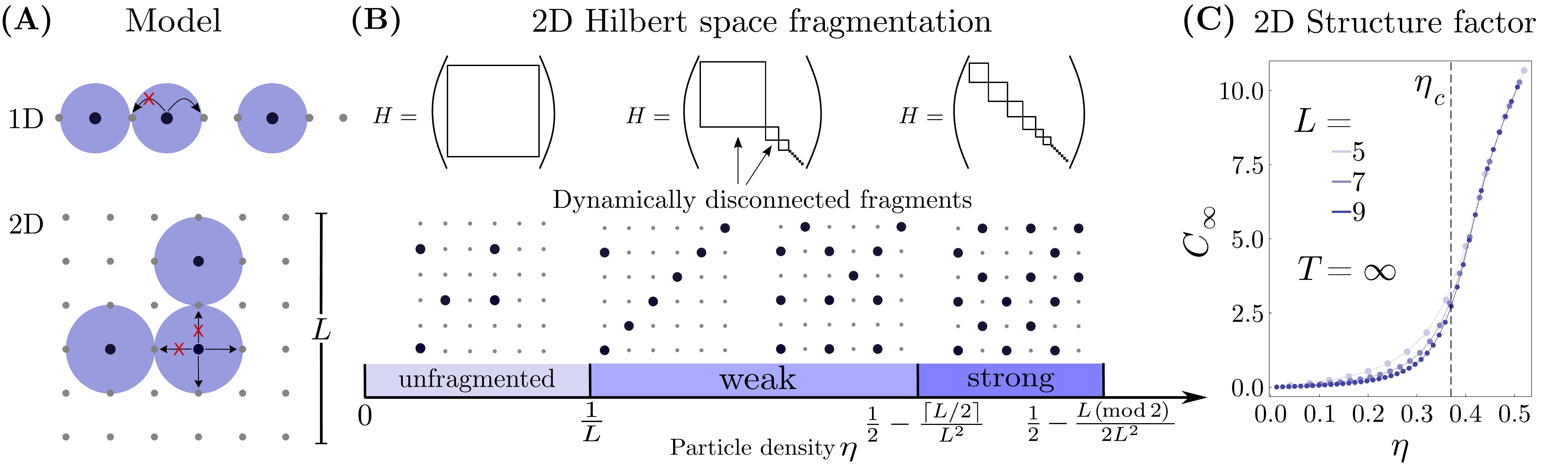}
    \caption{Hilbert-space fragmentation and 2D phase diagram. \textbf{(A)} An illustration of the considered hard-disk models on 1D chains and 2D square lattices. \textbf{(B)} Hilbert-space fragmentation in 2D as a function of particle density with representative configurations. \textbf{(C)} Infinite-temperature structure factor in 2D as a function of particle density $\eta$ for several system sizes $L \times L$.}
    \label{fig:FigureOne}
\end{figure*}

\textit{The quantum hard-disk model on a lattice.—} We model the quantum hard-disk problem as a system of hard-core bosons on a lattice with nearest-neighbor hopping:
\begin{equation}
    H = J\sum_{\langle i, j \rangle}  P_i \left(   {a}_i^{\dagger}  a_j +  a_j^{\dagger}  a_i \right) P_j.
    \label{eq:HamDressed}
\end{equation}
\noindent
Here, $a_i^{\dagger}$ and $a_i$ are the corresponding creation and annihilation operators for a hard-core boson on lattice site $i=1,\dots L^d$, respectively, with $L^d$ denoting the total number of lattice sites, and $d$ is the dimension of the lattice.
We set the lattice constant $a=1$ in the following.
The excluded volume due to the hard disk we include through the projection operators $P_i =  \prod_{ j \ni |\vec r_i - \vec r_j| = 1  }  (1 - n_j)/2$, 
that prevents particles from occupying nearest-neighboring sites (see Fig.\,\ref{fig:FigureOne}\textbf{A} for an illustration), and $n_i \equiv  a_i^{\dagger} a_i$.
Notice, that it is straightforward to adjust the disk radius by adjusting these projection operators to include more lattice sites.
%
%In the present work, we focus on disks with a radius of one lattice site.
%
The above Hamiltonian is equivalent to a model of spin-1/2's upon identifying $a_i \mapsto S_i^-$, $a_i^\dag \mapsto S_i^+$, and $n_i \mapsto S_i^z+1/2$ \cite{kwan2023minimal}, which links directly to a system of Rydberg atoms, see below for a more detailed discussion on the potential experimental realization.
We will consider open boundary conditions for convenience, which is also motivated by potential experimental scenarios in quantum simulators.

\textit{Hilbert-space fragmentation.—}A general feature of strong local constraints is the fragmentation of the Hilbert space into kinetically disconnected regions~\cite{Pai2019,Sala2020,Rakovszky2020,Khemani2020,PhysRevX.12.011050,Smith2017,Smith2017_2,Brenes2018,Russomanno2020,Karpov2021,Chakraborty2022,Chakraborty2022_2}, which can have drastic consequences on dynamical properties such as the emergence of quantum many-body scars~\cite{Bernien_2017,Turner2018,Michailidis2020,Serbyn2021} or disorder-free localization in gauge theories~\cite{Smith2017,Smith2017_2,Brenes2018,Russomanno2020,Karpov2021,Chakraborty2022,Chakraborty2022_2}.
In the thermodynamic limit one can distinguish weakly ($\mathcal{N}_{\textrm{max}}/\mathcal{N} \rightarrow 1$)  and strongly fragmented regimes ($\mathcal{N}_{\textrm{max}}/\mathcal{N} \rightarrow 0$)~\cite{PhysRevX.12.011050}, where $\mathcal{N}_{\textrm{max}}$ denotes the size of the largest fragment  and $\mathcal{N}$ the Hilbert-space dimension.

In the 1D case, we find that there is no fragmentation except at the maximum particle density $\eta = 1/2$.
However, in 2D, the system exhibits both weak and strong fragmentation depending on the particle density $\eta=M/L^d$ with $M$ the total number of particles, see Ref.~\cite{kwan2023minimal}.
Now we determine the precise threshold densities, illustrated in Fig.\,\ref{fig:FigureOne}\,\textbf{B}.
A configuration with occupied sites on the diagonal (termed a snake) is a static object due to the excluded volume and creates a barrier confining particles to the two regions. 
Consequently, we obtain Hilbert-space fragmentation whenever the particle density $\eta\geq 1/L$ or equivalently whenever the particle number $M\geq L$ becomes larger than the linear extent $L$ of the system.
For not too high densities, this leads to a weakly fragmented situation with the Hilbert space breaking up into several small fragments containing the states with snakes and a large fragment without snakes.
However, when $\eta \ge (1/2) - \lceil L/2 \rceil/L^2$ (with $\lceil L/2 \rceil$ denoting the ceiling operation), each allowed configuration must contain at least one snake or completely filled diagonals adjacent to the main diagonal, preventing large fragments and leading to strong fragmentation.
Fragmentation also facilitates the numerical solution by means of exact diagonalization. 
To simulate the quantum dynamics, instead of addressing the full Hamiltonian, we only solve for the relevant Hamiltonian blocks.
For the dynamics, we employ a Lanczos algorithm with 7 Krylov vectors at each time step.
We choose the time steps as large as possible while still achieving convergence.
In a subsequent section, we will compare the resulting quantum dynamics to a classical analog.
Viewing quantum dynamics as a quantum walk of hard disks on the lattice, we perform its natural classical counterpart—the random walk.

\textit{Static properties.—}Let us first discuss the static equilibrium properties of the quantum hard-disk problem on a lattice. Since we will be mostly interested in the dynamics at high energies, we focus in the following on the case of infinite temperature $T=\infty$. Classically, it is known that there is a dilute-crystalline phase transition in 2D~\cite{Fernandes_2007}. Since the crystalline phase can be characterized by the structure factor, here we also take it as the natural order parameter: $ C_{\infty}(\eta) = N^{-1} (2\pi/L^d)^{2}\sum_{i,j}^{} e^{i \vec \pi \cdot (\vec r_i - \vec r_j)} \Tr \left(n_i \, n_j\right)$.
%\begin{equation}
%    C_{\infty}(\eta) = \frac{1}{N} \left(\frac{2\pi}{L^d}\right)^{2}\sum_{i,j}^{} e^{i \vec \pi \cdot (\vec r_i - \vec r_j)} \Tr \left(n_i \, n_j\right).
%    \label{eq:SF}
%\end{equation}
\noindent
The trace is taken for the Hamiltonian blocks with particle density $\eta$, $N$ denotes the number of states in the block, and $\vec \pi$ is $\pi$ for 1D and equal to $(\pi, \pi)$ for 2D. Clearly, there is no difference to the classical case because the quantum trace becomes equal to the sum over all classical configurations and therefore the static phase diagrams are identical.

In 1D, this implies that there is crystalline order only at $\eta=1/2$. For 2D, the system exhibits a critical particle density $\eta_c = 0.37$ with a crystalline phase for $\eta>\eta_c$~\cite{Fernandes_2007}.
Notice that the crystalline phase transition is located in the weakly fragmented regime detached from the fragmentation thresholds.
The finite-size behavior of the structure factor we display in Fig.\,\ref{fig:FigureOne}\textbf{C}.
In the next step, we aim to study the dynamical properties, which will exhibit distinct quantum features.

\begin{figure}
    \centering
    \includegraphics[width=\linewidth]{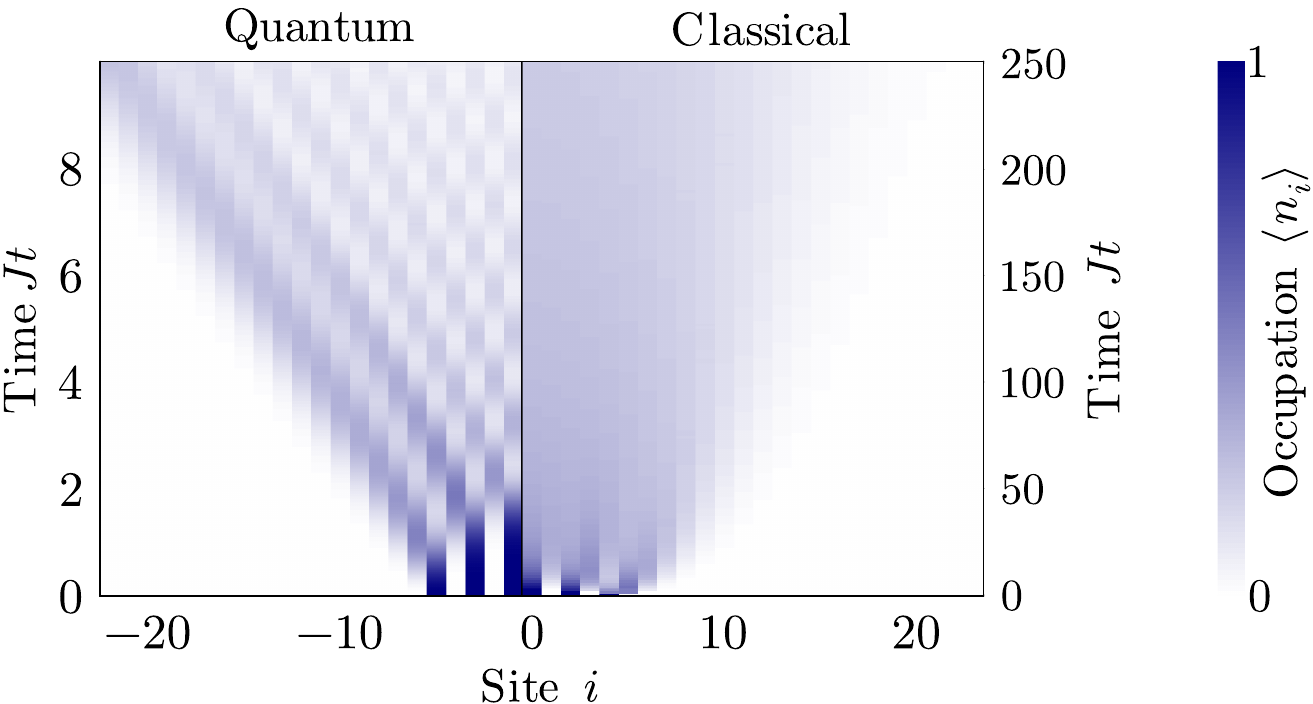}
    \caption{Crystal melting in 1D comparing quantum versus classical dynamics. The average on-site occupation $\langle n_i(t) \rangle$ is displayed as a function of time $Jt$ for an initial finite-size crystal with $M=5$ particles. Due to the symmetry of the spreading about the center of the crystal, we show the quantum and classical melting on the left and right halves, respectively. The classical time as a function of the number of steps taken $N_{\textrm{steps}}$ is given by, $Jt = 2\pi N_{\textrm{steps}}/M$.}
    \label{fig:Spread}
\end{figure}

\textit{Crystal melting in 1D.—}For the 1D case, we study the melting of a finite-size crystal object of $M$ tightly packed particles embedded in an otherwise empty lattice.
In Fig.\,\ref{fig:Spread}, we display the numerically obtained data for the local occupations $\langle n_i(t) \rangle$.
We observe that the spreading of the occupations exhibits fundamentally different behaviors in the quantum and classical cases, which can be proven analytically by mappings to effective models where the excluded volume is eliminated.
In the quantum domain, this can be achieved by removing the site to the right of each particle in every basis configuration.
The unitary dynamics are then equivalent to the single-particle quantum walk that is known to be ballistic, which is consistent with the results in Fig.~\ref{fig:Spread} and with the Bethe-ansatz integrability of the 1D model~\cite{Alcaraz1999, Pozsgay2021}.
In the classical domain, one can perform a mapping from the hard-disk model to one where disks have no radius \cite{Grabsch_2023}, where, however, it is key to take into account that due to the hard-core interaction particles cannot pass through each other implying preservation of the order of particles.
The dynamics turn then out to be sub-diffusive according to the Kardar-Parisi-Zhang universality class~\cite{Grabsch_2023}.
As a key consequence, we find that the 1D quantum crystal melts much faster than the classical one.

\begin{figure*}
    \centering
    \includegraphics[width=\linewidth]{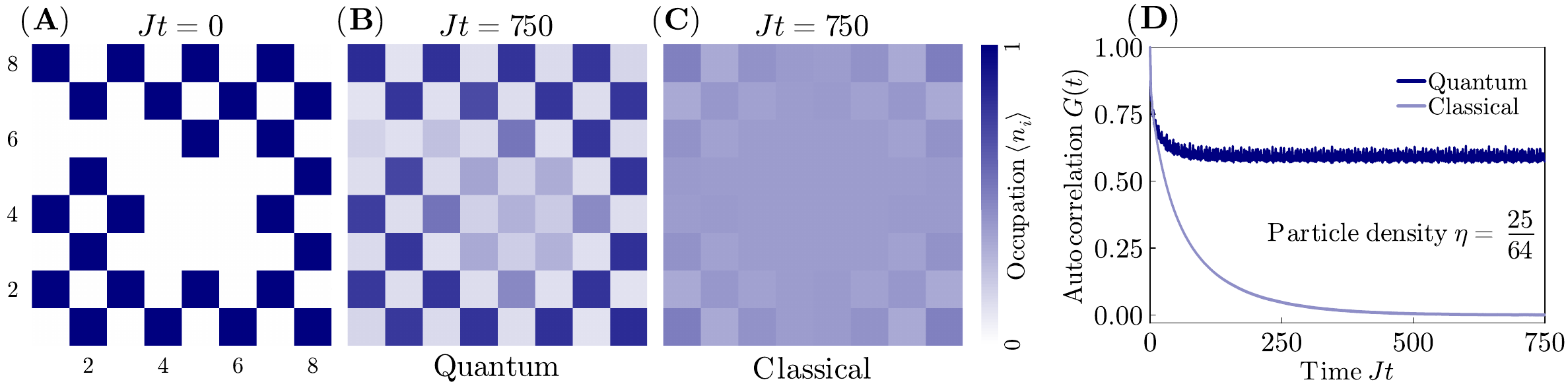}
    \caption{Defect dynamics on the square lattice comparing quantum and classical dynamics. (\textbf{A}) Initial defect configuration for a $8 \times 8$ lattice with density $\eta = 25/64$. (\textbf{B}) Long-time occupations $\langle n_i (t) \rangle$ at $Jt = 750$ for the quantum  (\textbf{C}) as well as the classical case. (\textbf{D}) Memory of the initial condition as measured through the autocorrelation function $G(t)$ for both the classical and quantum dynamics as a function of time $Jt$. The classical average is computed for $10^6$ trajectories.}
    \label{fig:Defects}
\end{figure*}

\textit{Defect dynamics in 2D.—}
We will now explore the nonequilibrium real-time evolution of crystal defects in 2D and we will show that their dynamics exhibit genuine quantum features.
For that purpose, we prepare the system in a specific many-body configuration and monitor the subsequent dynamics, see Fig.~\ref{fig:Defects}.
In the beginning, let us briefly discuss the limits of weak and high particle densities.
Clearly, preparing a configuration with density $\eta<1/L$ in the non-fragmented regime the excluded volume becomes irrelevant and the dynamics of the particles becomes essentially equivalent to a simple tight-binding problem and any initial configuration is smeared out uniformly.
In the opposite regime $\eta \ge (1/2) - \lceil L/2 \rceil/L^2$, the strong fragmentation of Hilbert space naturally leads to nonergodic behavior leaving any crystal intact.
Most importantly, in the regime of weak fragmentation, the dynamics yields unexpected behavior, see Fig.~\ref{fig:Defects}.
For the initial configuration in Fig.~\ref{fig:Defects}(A), we observe that the detailed crystal structure remains stable under quantum evolution, whereas classically the crystal structure is completely washed out.
Accordingly, the quantum system retains a memory as opposed to the classical counterpart calling for the importance of quantum interference effects, a key aspect absent in the random walk. 
As we will discuss, this is a generic behavior not fine-tuned to this particular initial condition.
We quantify the memory of the crystal pattern through:
\begin{equation}
    G(t) = \frac{1}{L^2}\sum_{i=1}^{L^2} \left\langle \left(2n_i(t) - 1\right) \left(2n_i(0) - 1\right)\right\rangle - G^*~.
\end{equation}
\noindent
Here, $\langle \dots \rangle = \langle \psi | \dots | \psi \rangle$ represents the expectation value with $|\psi\rangle$ the initial condition for the quantum case and an average over trajectories for the classical one.
A constant $G^* = \left(2\eta - 1\right)^2$ is subtracted such that $G(t) \rightarrow 0$ when the crystal melts and $\langle n_i(t)\rangle \to \eta$.

In Fig.~\ref{fig:Defects}(D), we display $G(t)$ as a function of time $t$ for the initial condition in Fig.~\ref{fig:Defects}(A).
For the quantum evolution, $G(t)$ saturates to a nonzero value, while in the classical domain $G(t)$ decays.
Notice that classically $G(t\rightarrow\infty)$ can retain a small non-zero value due to finite-size effects vanishing in the thermodynamic limit~\cite{Supp}.

\begin{figure*}
    \centering
    \includegraphics[width=\linewidth]{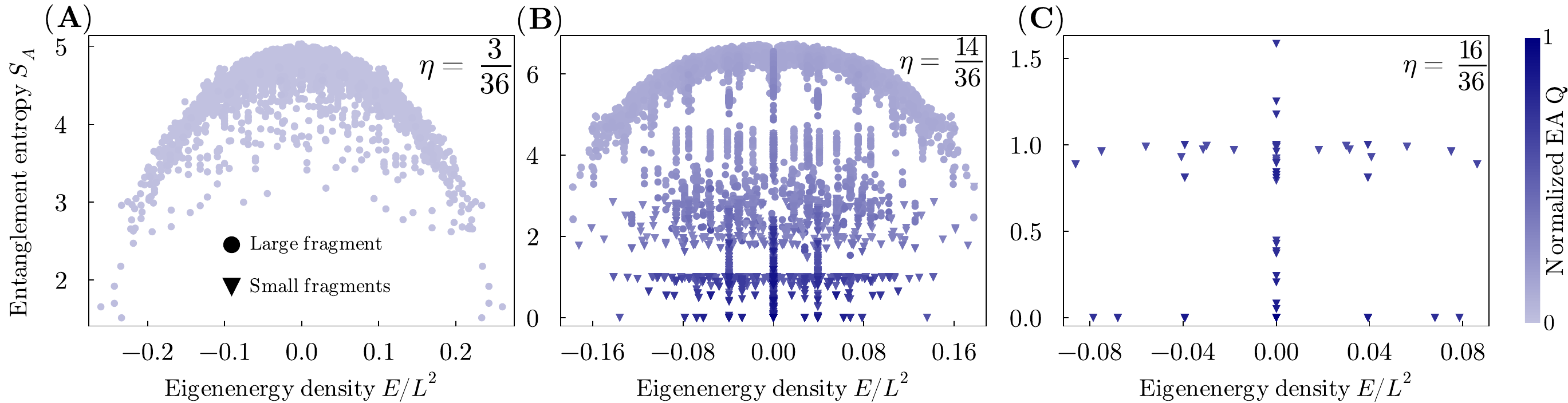}
    \caption{Quantum many-body scars in the 2D quantum hard-disk model. Bipartite entanglement entropy $S_A$ as a function of eigenstate energy density colored by the normalized Edwards-Anderson order parameter $Q$ for a $6 \times 6$ lattice. A circle denotes eigenstates from the largest Hilbert-space fragment whereas triangles indicate eigenstates from the smaller ones. (\textbf{A}) In the non-fragmented region, there is only a large ergodic component.
    (\textbf{B}) In the weak fragmentation region, there are towers of eigenstates with low entanglement entropy and high normalized Edwards-Anderson order parameter $Q$ signaling quantum many-body scars. (\textbf{C}) For high densities in the regime of strong fragmentation all eigenstates lie in small Hilbert-space fragments.}
    \label{fig:EE}
\end{figure*}

The defect dynamics in Fig.~\ref{fig:Defects} highlights the crucial role of quantum interference effects leading to nonergodic behavior.
In the following, we provide compelling numerical evidence that this originates from the presence of quantum many-body scars in the spectrum.
For that purpose, we will study the eigenstate properties through the bipartite entanglement entropy:
\begin{equation}
    S_A = -\Tr \bigg[\Tr_{\bar{A}} \big(|\Psi \rangle \langle \Psi |\big) \log \left[ \Tr_{\bar{A}} \big(|\Psi \rangle \langle \Psi |\big)\right]\bigg].
\end{equation}
\noindent
Here, $|\Psi\rangle$ denotes an eigenstate, $A$ the subsystem, chosen as the bottom half of the 2D lattice, and $\bar{A}$ is the complement of $A$.
Additionally, we aim to quantify how close a given eigenstate is to a single many-body configuration, such as the one for the dynamics shown in Fig.~\ref{fig:Defects}.
This can be achieved by means of an Edwards-Anderson (EA) order parameter $ Q_{\textrm{EA}} = L^{-4} \sum_{i,j}^{} |\langle \Psi |(2 n_i - 1) (2 n_j - 1 ) |\Psi \rangle |^2$.
In order to quantitatively compare the EA order parameter for different particle densities $\eta$ we further introduce a normalized version:
\begin{equation}
    Q = \frac{Q_{\textrm{EA}} - \left(2\eta - 1\right)^4}{1 - \left(2\eta - 1\right)^4}~,
\end{equation}
\noindent
so that $Q=1$ when $|\Psi\rangle$ is a single configuration and $Q \rightarrow 0$ when $|\Psi\rangle$ represents a superposition over all basis states.

In Fig.\,\ref{fig:EE} we display $S_A$ for all eigenstates as a function of energy density $E/L^2$ for three different densities $\eta$ and associated levels of fragmentation.
We include the normalized EA order parameter $Q$ by coloring the data points.

In the non-fragmented domain, see Fig.~\ref{fig:EE}(A), we observe a dome for $S_A$ consistent with ergodic behavior, further supported by the uniform small values of $Q$.
In the strong fragmentation regime, see Fig.~\ref{fig:EE}(C), there is no ergodic component but rather only small fragments yielding small values of $S_A$ and high values of $Q$. 
Consequently, the eigenstates represent almost individual many-body configurations.

Concerning the stability of defects in Fig.~\ref{fig:Defects}, the regime of most interest is the weakly fragmented region, see Fig.~\ref{fig:EE}(B).
There, one can identify a set of ergodic eigenstates.
On the other hand, there is a large number of eigenstates, which don't follow the conventional ergodic paradigm, which we identify as quantum many-body scars where $S_A$ and $Q$ significantly depart from the ergodic expectations.
Most importantly, these nonergodic eigenstates not only emerge from the small Hilbert-space fragments but also from the large one, where they arrange along towers at specific energy densities.
As these nonergodic eigenstates are mostly also associated with large values of $Q$, we conclude that there is a large number of many-body configurations, which are not thermalizing but rather retain their original structure over time.
This is exactly related to the stability of the crystal defects displayed in Fig.~\ref{fig:Defects} and we conclude that our initial configuration in Fig.~\ref{fig:Defects} was not fine-tuned.
%
%But rather there are many crystal defects, which leave the quantum crystal stable.
%
When it comes to the stability of our observed quantum many-body scars, it might happen that additional interactions yield dynamical heterogeneous behavior as has been studied for a related quasi-1D ladder system~\cite{PhysRevLett.121.040603}.
Notice that recent works in strong-coupling limits of the quantum Ising model have recognized also regimes of nonergodic behaviors for certain types of domain wall interfaces~\cite{Hart2022,Balducci2022,Balducci2023,Balducci2023_2}.

\textit{Conclusions.—}In this letter, we have studied the quantum hard-disk model on lattices, which as we have shown displays unique quantum features in its dynamics.
The considered hard-disk model exhibits a direct realization in Rydberg atomic systems.
The Rydberg blockade mechanism naturally implements the excluded volume, so that a Rydberg excitation (identified here with a hard-core bosonic particle) blocks for the presence of a second excitation within the blockade radius~\cite{Browaeys_2020}.
When considering Rydberg atoms on a lattice, it remains to adjust the lattice spacing between the atoms such that the Rydberg blockade radius is exactly such that it blocks the presence of particles on nearest-neighboring sites.
The typical microscopic interactions realized in Rydberg atomic systems are either of Ising character, leading to PXP-type models~\cite{Bernien_2017, Bluvstein2021}, or $XY$ character~\cite{Chen2023}.
Importantly, our hard-disk model exhibits particle-number conservation, which would be either implementable directly via $XY$-type interactions or via Ising-type models upon enforcing particle-number conservation through a strong longitudinal magnetic field for instance.

We envision a large range of further interesting prospective research questions for the future emerging from this work.
A natural extension would be to consider disks or excluded volumes with larger radii as well as different shapes.
Further, it might be interesting to consider how the dynamics are modified upon relaxing the hard-disk constraint to a soft one, such as for the soft-shoulder potentials realized naturally in Rydberg dressing~\cite{Zeiher2017}.

\textit{Data availability.—}The data to generate all figures in this letter is available in Zenodo \cite{Quantum_hard_disks_dataset}.

\textit{Acknowledgements.—}We thank Werner Krauth, Juan Garrahan, Alessio Lerose, and Tobias Wiener for fruitful discussions. V.D.N. and F.B.T. contributed equally to this work. This project has received funding from the European Research Council (ERC) under the European Union’s Horizon 2020 research and innovation programme (grant agreement No. 853443). 
This work was supported by the German Research Foundation DFG via project 499180199 (FOR 5522).


% \clearpage
\nocite{*}

\bibliography{References}% Produces the bibliography via BibTeX.
\clearpage

\end{document}

\title{Supplemental Material of “Quantum hard disks on a lattice"}

\author{Vighnesh Dattatraya Naik}
\author{Fabian Ballar Trigueros}
\author{Markus Heyl}
 % \email{}
\affiliation{%
 Theoretical Physics III, Center for Electronic Correlations and Magnetism,
Institute of Physics, University of Augsburg, 86135 Augsburg, Germany}%

\date{\today}% It is always \today, today,

\maketitle

%%%%%%%%%% Prefix a "S" to all equations, figures, tables and reset the counter %%%%%%%%%%
\setcounter{equation}{0}
\setcounter{figure}{0}
\setcounter{table}{0}
\setcounter{page}{1}
\makeatletter
\renewcommand{\theequation}{S\arabic{equation}}
\renewcommand{\thefigure}{S\arabic{figure}}
\renewcommand{\bibnumfmt}[1]{[S#1]}
\renewcommand{\citenumfont}[1]{S#1}
%%%%%%%%%% Prefix a "S" to all equations, figures, tables and reset the counter

\section{Bipartite entanglement entropy versus eigenstate energy density}
\begin{figure*}[h!]
    \centering
    \includegraphics[width=0.8\linewidth]{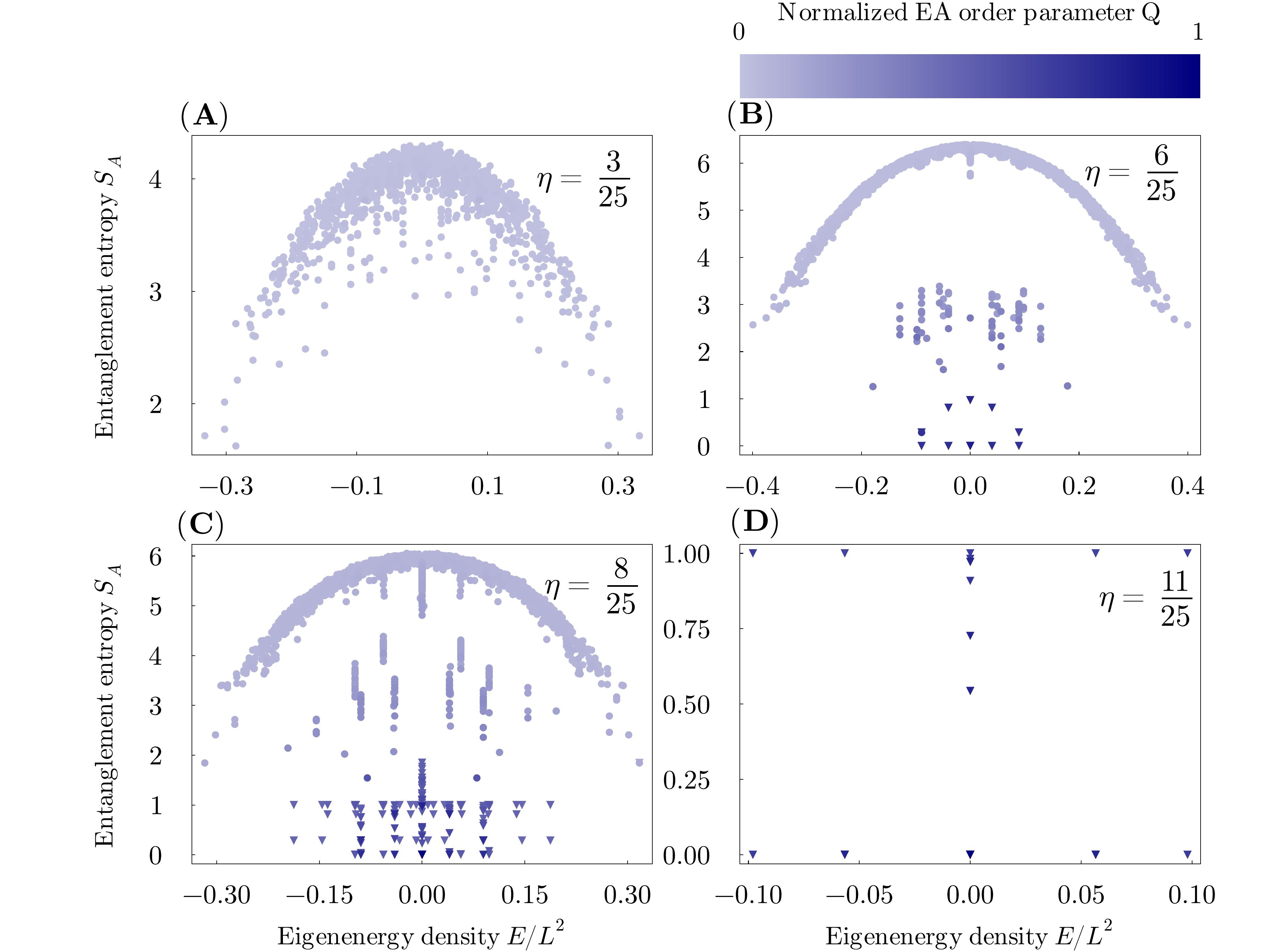}
    \caption{Bipartite entanglement entropy $S_A$ as a function of eigenstate energy density colored by Edwards-Anderson order parameter $Q_{\textrm{EA}}$ for $5 \times 5$ lattice. A circle denotes eigenstates from the largest fragment whereas triangles indicate eigenstates from the smaller fragments. (\textbf{A}) In the no fragmentation region, there is only a large ergodic component.
    (\textbf{B} and \textbf{C}) In the weak fragmentation region, there are eigenstates with low entanglement entropy lying outside of the ergodic dome signaling quantum many-body scars. (\textbf{D}) Due to the strong fragmentation, there is no ergodic component.}
    \label{fig:EE_L5}
\end{figure*}

In Fig.\,\ref{fig:EE_L5}, we show a plot analogous to that of Fig.\,4 of the main text for a $5 \times 5$ lattice. In the weak fragmentation region Fig.\,\ref{fig:EE_L5}(\textbf{B}-\textbf{C}), we observe that the scar states also appear and their number increases with density.  

\section{Different initial defects}

\begin{figure*}[h!]
    \centering
    \includegraphics[width=\linewidth]{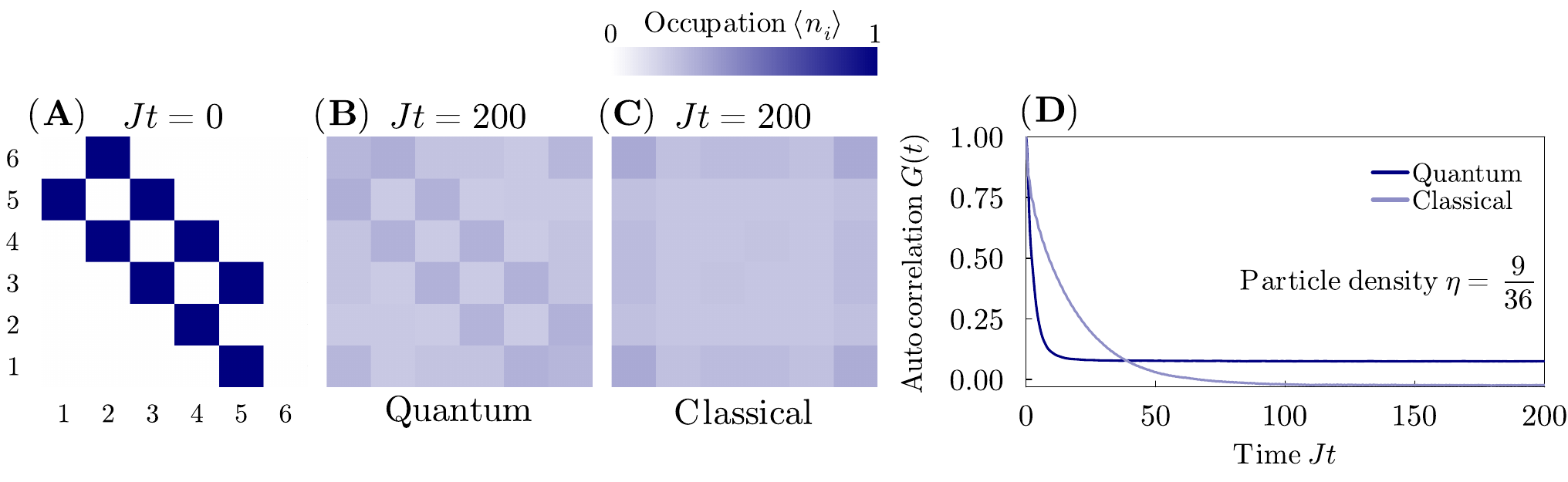}
    \caption{Defect dynamics for $6 \times 6$ lattice with density $\eta = 9/36$. (\textbf{A}) Initial defect configuration. (\textbf{B}) $\langle n_i \rangle$ at $Jt = 200$ for the quantum case. (\textbf{C}) $\langle n_i \rangle$ at $Jt = 200$ for the classical case. (\textbf{D}) Autocorrelation function for both the classical and quantum cases as a function of time. The classical average is computed for $10^5$ trajectories.}
    \label{fig:Defects_init3}
\end{figure*}

In this section, we present results for the comparison between the 2D classical and quantum dynamics for different system sizes and initial defect configurations in the weak fragmentation region. In Fig.\,\ref{fig:Defects_init3} we show the dynamics for an initial crystal in a $6 \times 6$ lattice with a very low density. Interestingly, even at this low density, we see evidence of quantum effects coming from the overlap of the initial state with scar states.\\

\begin{figure*}
    \centering
    \includegraphics[width=\linewidth]{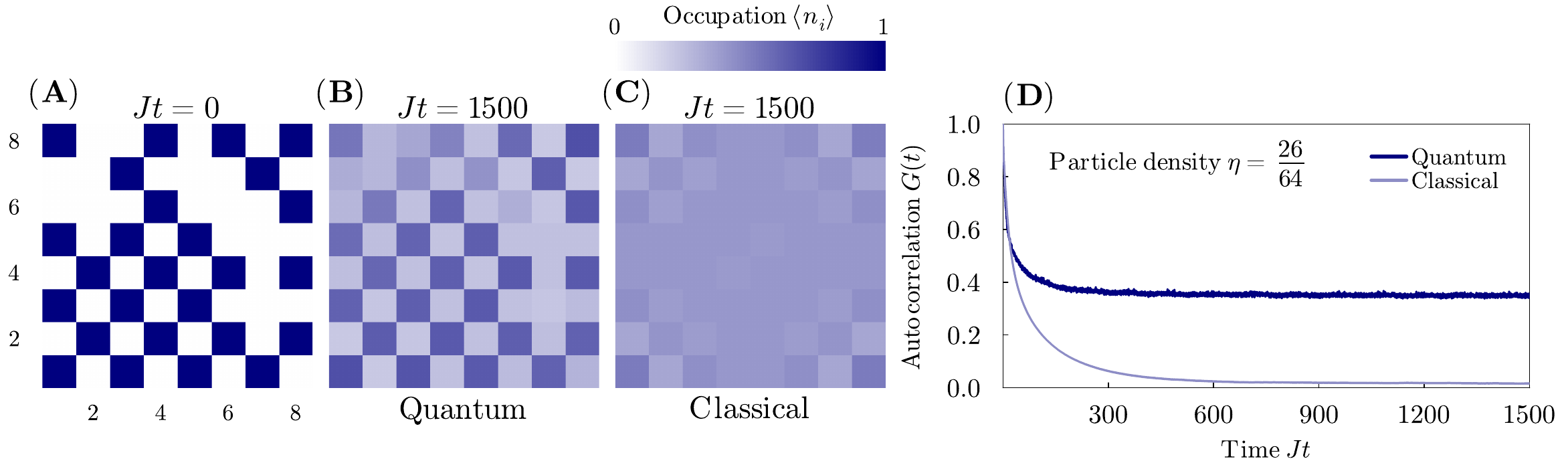}
    \caption{Defect dynamics for $8 \times 8$ lattice with density $\eta = 26/64$. (\textbf{A}) Initial defect configuration. (\textbf{B}) $\langle n_i \rangle$ at $Jt = 1500$ for the quantum case. (\textbf{C}) $\langle n_i \rangle$ at $Jt = 1500$ for the classical case. (\textbf{D}) Autocorrelation function for both the classical and quantum cases as a function of time. The classical average is computed for $10^6$ trajectories.}
    \label{fig:Defects_init5}
\end{figure*}

\begin{figure*}[h!]
    \centering
    \includegraphics[width=\linewidth]{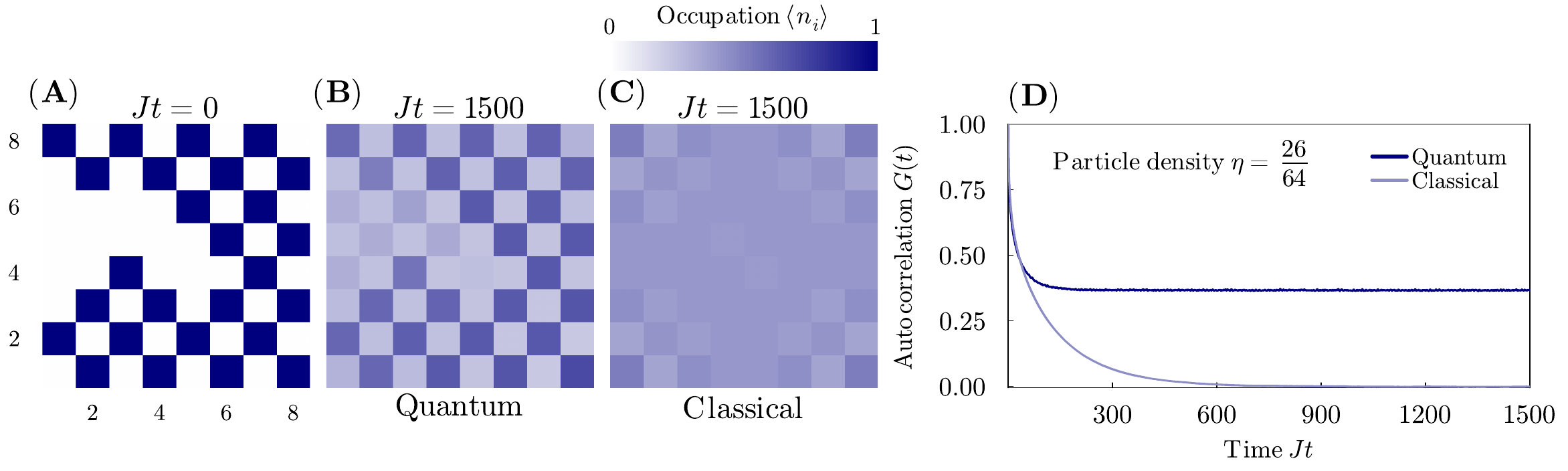}
    \caption{Defect dynamics for $8 \times 8$ lattice with density $\eta = 26/64$. (\textbf{A}) Initial defect configuration. (\textbf{B}) $\langle n_i \rangle$ at $Jt = 1500$ for the quantum case. (\textbf{C}) $\langle n_i \rangle$ at $Jt = 1500$ for the classical case. (\textbf{D}) Autocorrelation function for both the classical and quantum cases as a function of time.  The classical average is computed for $10^6$ trajectories.}
    \label{fig:Defects_init2}
\end{figure*}

\begin{figure*}[h!]
    \centering
    \includegraphics[width=\linewidth]{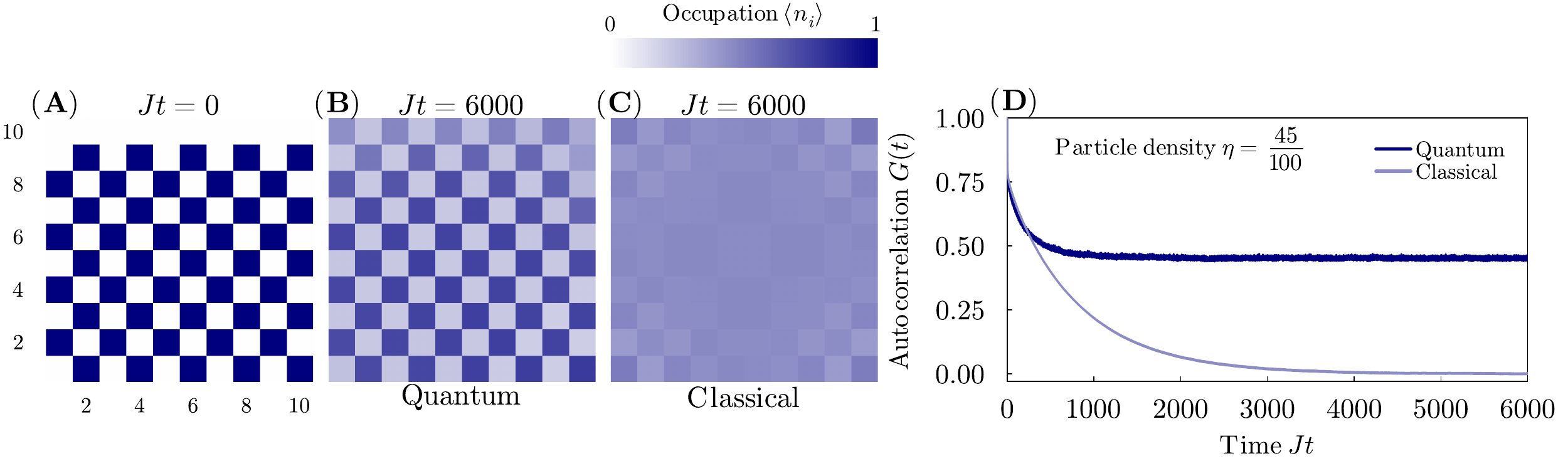}
    \caption{Defect dynamics for $10 \times 10$ lattice with density $\eta = 45/100$. (\textbf{A}) Initial defect configuration. (\textbf{B}) $\langle n_i \rangle$ at $Jt = 6000$ for the quantum case. (\textbf{C}) $\langle n_i \rangle$ at $Jt = 6000$ for the classical case. (\textbf{D}) Autocorrelation function for both the classical and quantum cases as a function of time. The classical average is computed for $10^6$ trajectories.}
    \label{fig:Defects_init4}
\end{figure*}

To give further examples of the same system size as the one presented in Fig.\,3 of the main text, we show different initial defect configurations that do not melt (see in Fig.\,\ref{fig:Defects_init5} and \ref{fig:Defects_init2}). Lastly, in Fig.\,\ref{fig:Defects_init4}, we show a larger system size of $10 \times 10$ with a high density where the system still retains a large memory of the initial state whereas it completely melts for the classical case.

\section{Finite size effects in 2D classical dynamics}

In this section, we want to discuss the boundary effects that are present in the 2D classical dynamics. Concretely, the crystal structure that appears near the boundaries in the final occupation. Leading $G(t)$ to not take a value of exactly $0$ even though all memory of the initial state is lost, as all initial states evolve to the same final occupation. To show that these boundary effects are negligible in the thermodynamic limit we measure the mean deviation of the final occupation from a homogeneous occupation fixed by the particle density $\eta$ defined as 

\begin{equation}
    \delta\eta = \frac{1}{L^2}\, \sum_i |\langle n_i\rangle - \eta|.
\end{equation}

\begin{figure*}[h!]
    \centering
    \includegraphics[width=0.5\linewidth]{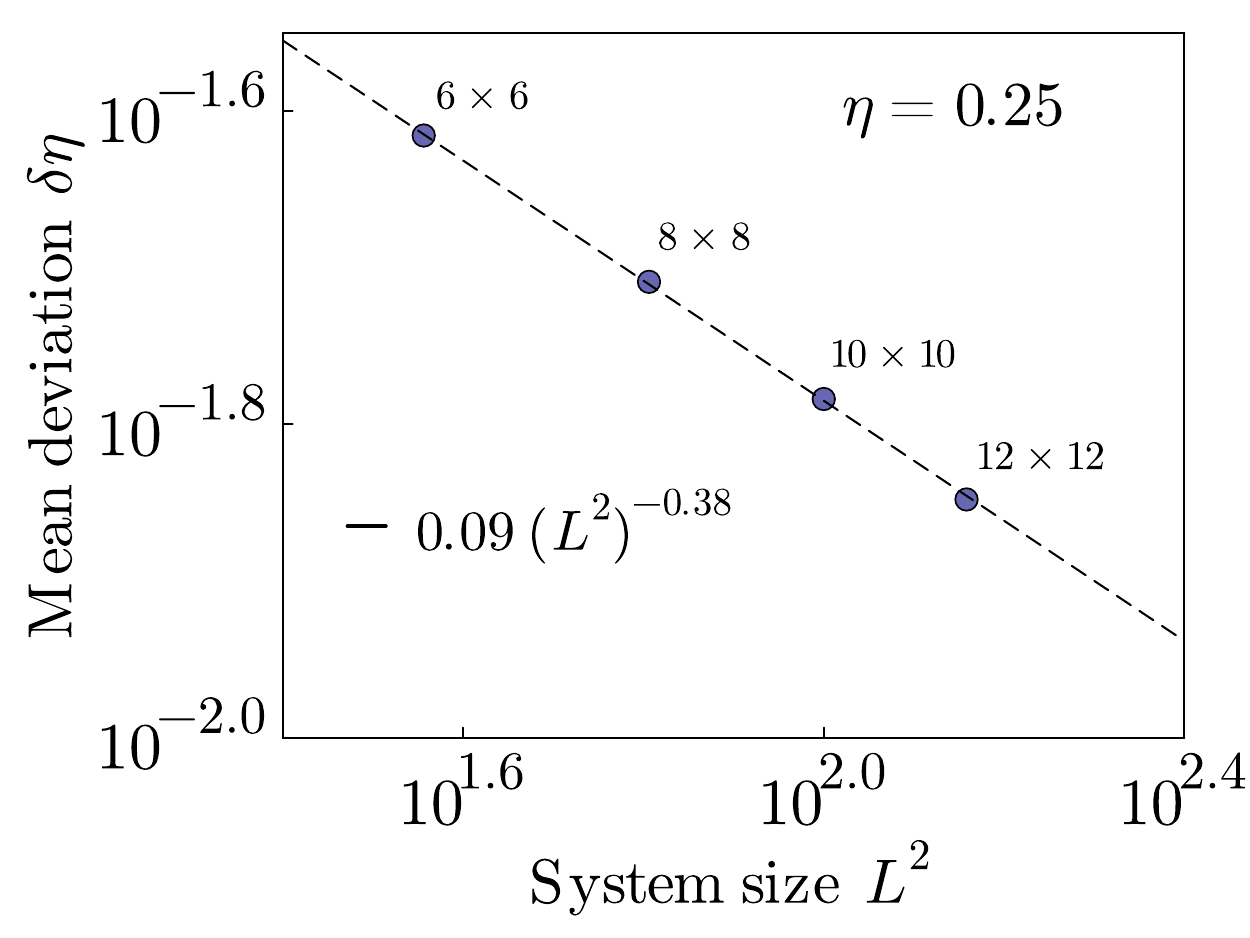}
    \caption{Scaling of mean deviation of the final occupation from a homogeneous occupation as a function of system size at a fixed density.}
    \label{fig:Scaling}
\end{figure*}

In Fig.\,\ref{fig:Scaling}, we show the mean deviation as a function of system size with the initial state being maximally packed in the bottom half of the lattice. We find that the deviation decreases as a power law with increasing system size. Showing that, in the thermodynamic limit the autocorrelation $G(t)$ will become $0$.